\batchmode

\documentclass[12pt]{article}

\usepackage[hmargin=1in,vmargin=1in]{geometry}

\usepackage{pslatex}
\usepackage{natbib}
\usepackage{fancyhdr}
\usepackage{graphicx}
\usepackage{xcolor}

\bibpunct{}{}{}{s}{}{} 

\title{\bf
An Internet Approach\\
for Engineering Student Exercises
}

\date{}

\author{Richard Perry\\
Department of Electrical and Computer Engineering\\
Villanova University, Villanova, PA 19085
}

\begin{document}

\setlength{\parindent}{0.0in}

\maketitle 

\thispagestyle{fancy}

\begin{abstract}

\vspace{-22pt}

\noindent
An approach for engineering student exercises using the Internet
is described.  In this approach, for a given exercise, each student
receives the same problem, but with different data.  
The exercise content can be static or dynamic,
and the dynamic form can be timeless or real--time.
The implementation provides immediate feedback to the students, letting them
know if their submitted answers are correct.  
Student results for each exercise are recorded in log files which are
available to the instructor.
Example exercises from engineering computer security and cryptography courses
are presented.

\end{abstract}
\newpage

Introduction

\vspace{11pt}

An approach for engineering student exercises using the Internet
is described.  In this approach, for a given exercise, each student
receives the same problem, but with different data.
Alternatively,
each student could receive different or slightly different problems
with the same or different data.  

\vspace{11pt}

The exercise content can be static or dynamic.  
In the static form, each time a student 
accesses an exercise, the same data is presented.
This allows students to work offline on problems and return
later to submit their solutions.  
In the dynamic form, 
each time a student accesses an exercise, different data is presented.
The data
is generated pseudo--randomly, based on the student UserID, so it
can be reproduced for the static exercises.

\vspace{11pt}

The dynamic form can be timeless or real--time.  For a timeless dynamic
exercise, students can work offline on the problem as long as the original
data web page is preserved in their browser or saved in a file.
For real--time dynamic exercises, students must submit their answers
within a small time window, e.g.\ 60 seconds.
The real--time exercises are implemented using a custom server process
running on an Internet site; the other types of exercises are implemented
using a standard web server environment.

\vspace{11pt}

The implementation provides immediate feedback to the students, letting them
know if their submitted answers are correct.  
For a multi--part exercise,
which requires a sequence of answers, this allows the students to complete
the exercise part--by--part, moving on to successive sections as each
part in the sequence is completed correctly.  
For correctly completing an exercise, students may be given a ``reward''.
Examples of rewards are: an opportunity to try a harder exercise for extra credit;
a random interesting adage from the Unix fortune utility; or simply
a congratulatory statement.

\vspace{11pt}

Student results for each exercise are recorded in log files which are
available to the instructor.  The results can be easily processed in an
automated fashion for grading.  The log files are also useful for analyzing
the behavior of students by looking at the time they started working on an
exercise (e.g.\ a week before it was due or an hour before it was due) and
how many incorrect attempts they made before submitting the correct answers.
This can help the instructor identify students who are performing extra well,
or who may need extra help.

\vspace{11pt}

Example exercises from engineering computer security and cryptography courses
will be presented, including a man--in--the--middle scenario and an exercise
in secure authentication and confidentiality over an insecure channel.

\vspace{22pt}
Static Examples
\vspace{11pt}

\begin{figure}[ht]
\begin{center}
\fbox{\includegraphics[scale=0.5,angle=0]{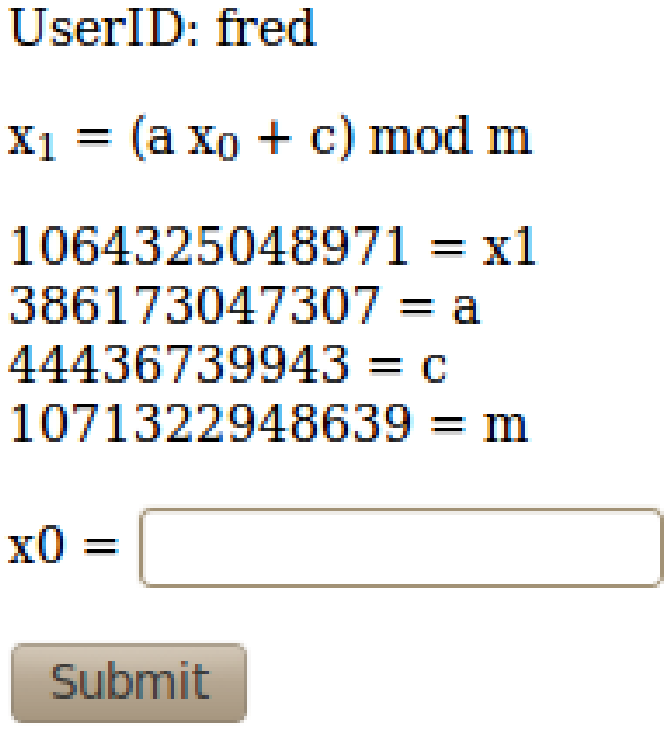}}
\caption{Static Example with a Single Answer}
\label{fig:seed}
\end{center}
\end{figure}

Figure \ref{fig:seed} shows an example where the student must solve a
modular equation for a single unknown.  The equation represents a linear
congruential pseudo--random number generator.
Here the student has submitted an incorrect answer:

\begin{verbatim}
    UserID: fred

    Your answer for x0 is wrong.
\end{verbatim}

and here the student has submitted the correct answer:

\begin{verbatim}
    UserID: fred

    Your answer for x0 is correct.
\end{verbatim}

\begin{figure}[ht]
\begin{center}
\fbox{\includegraphics[scale=0.5,angle=0]{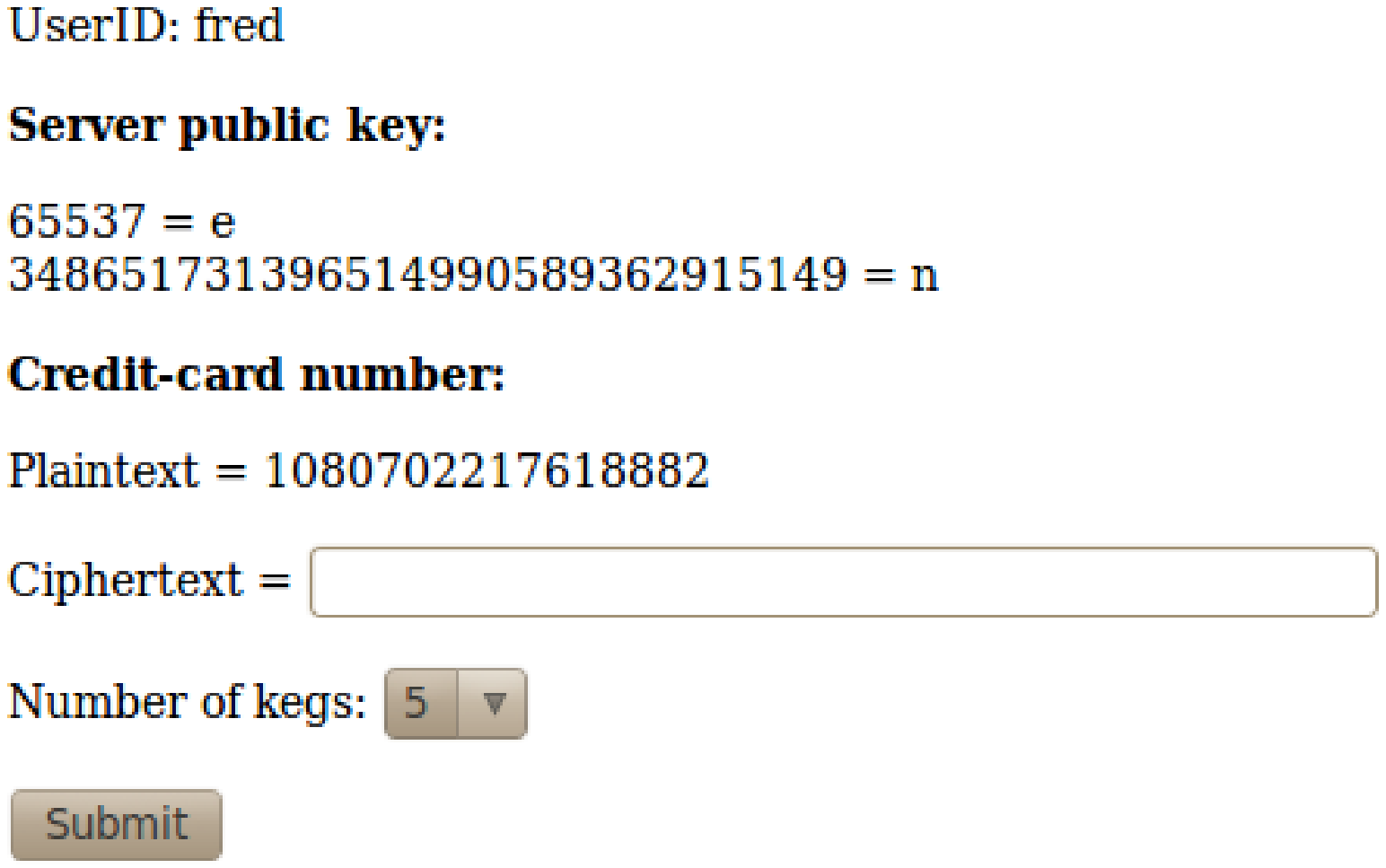}}
\caption{Static Example using RSA to Digitally Sign a Credit Card Number}
\label{fig:milk}
\end{center}
\end{figure}

\vspace{11pt}

Figure \ref{fig:milk} shows another static example
using RSA to digitally sign a ``credit--card'' number for a fake online
transaction, with an option to specify how many kegs of ``milk'' to order.
The option is ignored when checking the results, but it makes the exercise
a little more fun for the students.  An example of correct results:

\begin{verbatim}
    UserID: fred

    Your credit-card number is valid.

    Your milk order will be shipped today!
\end{verbatim}

\begin{figure}[ht]
\begin{center}
\fbox{\includegraphics[scale=0.5,angle=0]{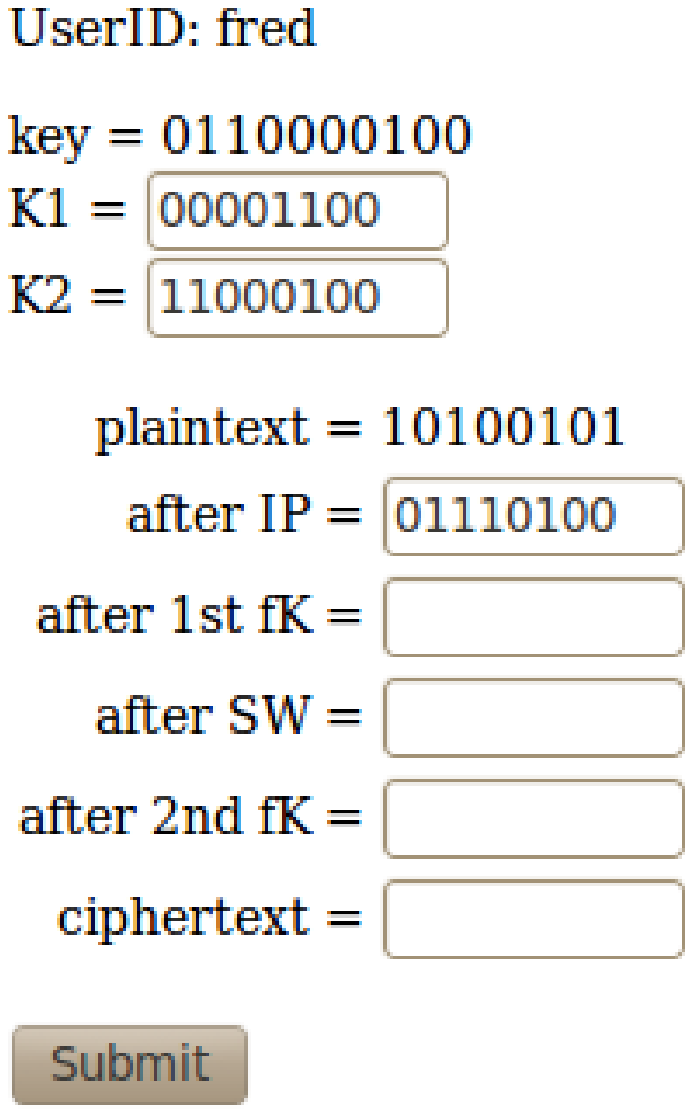}}
\caption{Static Multi--part Example}
\label{fig:sdesp}
\end{center}
\end{figure}

A multi--part example, requiring a sequence of answers, is shown in
Fig.\ \ref{fig:sdesp}.
In this exercise the student must perform encryption
and show intermediate results using a simplified form of the Data Encryption Standard.
An incorrect calculation by the
student for one part would cause subsequent parts to be incorrect.
By submitting partial results, the student is able to complete
the exercise part--by--part, moving on to successive parts as each
step in the sequence is completed correctly:

\begin{verbatim}
    UserID: fred

    Your answer for K1 is correct.
    Your answer for K2 is correct.
    Your answer for IP is correct.
    Your answer for fK1 is wrong.
    Your answer for SW is wrong.
    Your answer for fK2 is wrong.
    Your answer for c is wrong.

    You have 3 out of 7 parts correct.
\end{verbatim}

\begin{figure}[ht]
\begin{center}
\fbox{\includegraphics[scale=0.5,angle=0]{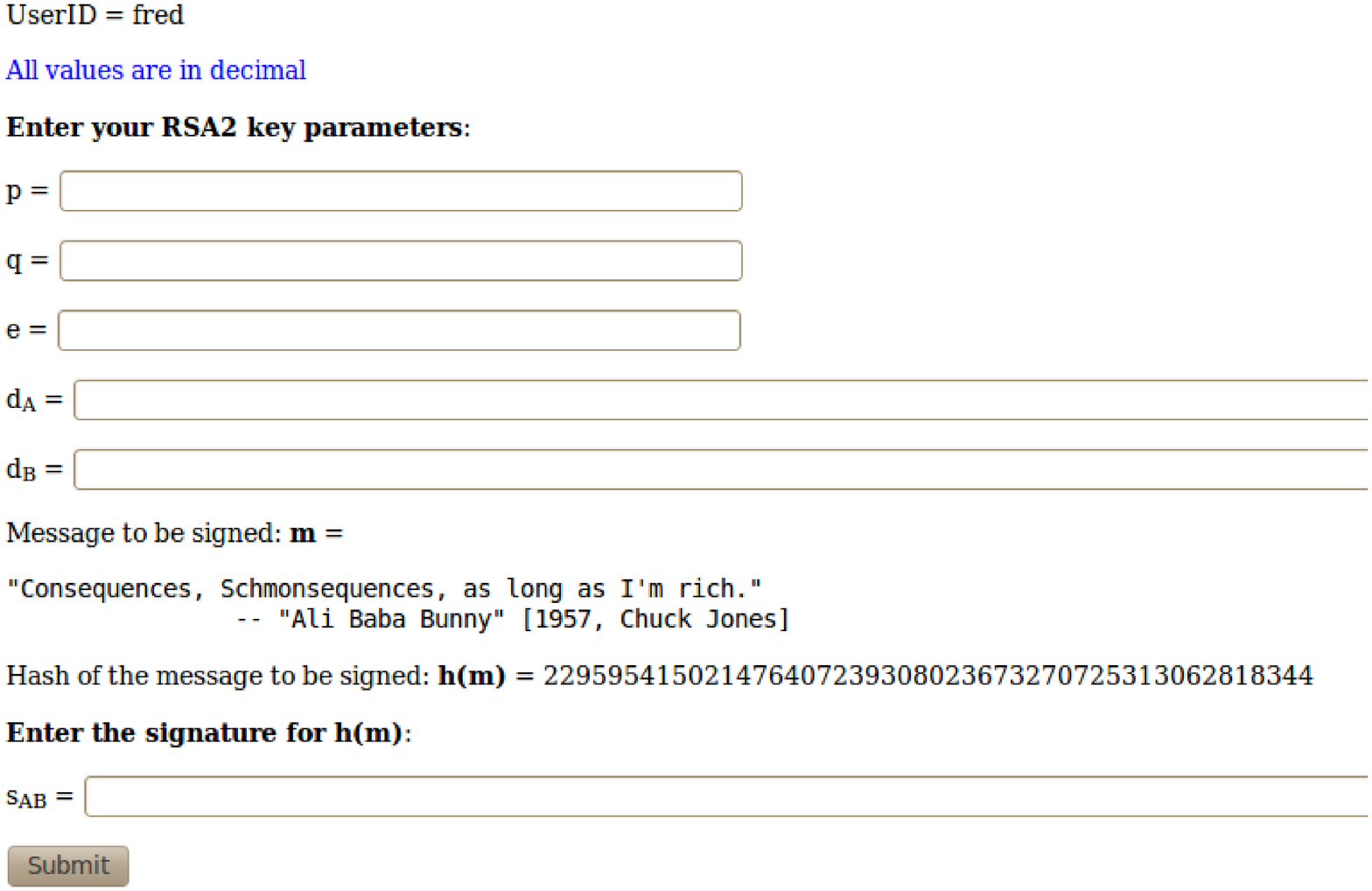}}
\caption{Two--user RSA Key Example}
\label{fig:RSA2}
\end{center}
\end{figure}

\vspace{11pt}

Figure \ref{fig:RSA2} shows an example
where most of the data is actually dynamically created by the student.
The student must use their UserID (treated as a base--36
number, plus 1 if it is even) as the public RSA exponent, but the other
values are left for the student to create on their own, within certain constraints.
The exercise involves designing a two--user split RSA key, and using the key
to produce a digital signature.  The message to be signed is
randomly generated using the Unix fortune utility, and changes
each time the student accesses the exercise, so this exercise is only partially static.

\vspace{11pt}

The process of checking the student results in this case
uses a conversational style, simulating what an instructor might do when
discussing student results in person.  At the end, if all six parts are correct,
the student is given positive feedback:

\begin{verbatim}
    UserID: fred
    
    e is correct, let's check p next:
    bitLength(p) == 128, good...
    p is prime, almost there for p...
    gcd(e,p-1) == 1, p is ok, let's check q next:
    q != p, that's a good start...
    bitLength(q) == 128, good...
    q is prime, almost there for q...
    gcd(e,q-1) == 1, q is ok, let's check d_A next:
    bitLength(d_A ) >= 240, good...
    gcd(d_A ,p-1) == 1, almost there for d_A ...
    gcd(d_A ,q-1) == 1, d_A is ok, let's check d_B next:
    d_B != 1, that's a good start...
    e*d_A *d_B == 1, Brilliant!
    h(m) is valid, checking signature:
    s_AB is valid.

    You have 6 out of 6 parts correct. You are the master of RSA2!
\end{verbatim}

\newpage
% \vspace{22pt}
Timeless Dynamic Examples
\vspace{11pt}

Each time a student accesses a dynamic exercise, different data is presented.
For a timeless dynamic exercise,
students can work offline on the problem as long the original
data web page is preserved in their browser or saved in a file.

\begin{figure}[ht]
\begin{center}
\fbox{\includegraphics[scale=0.5,angle=0]{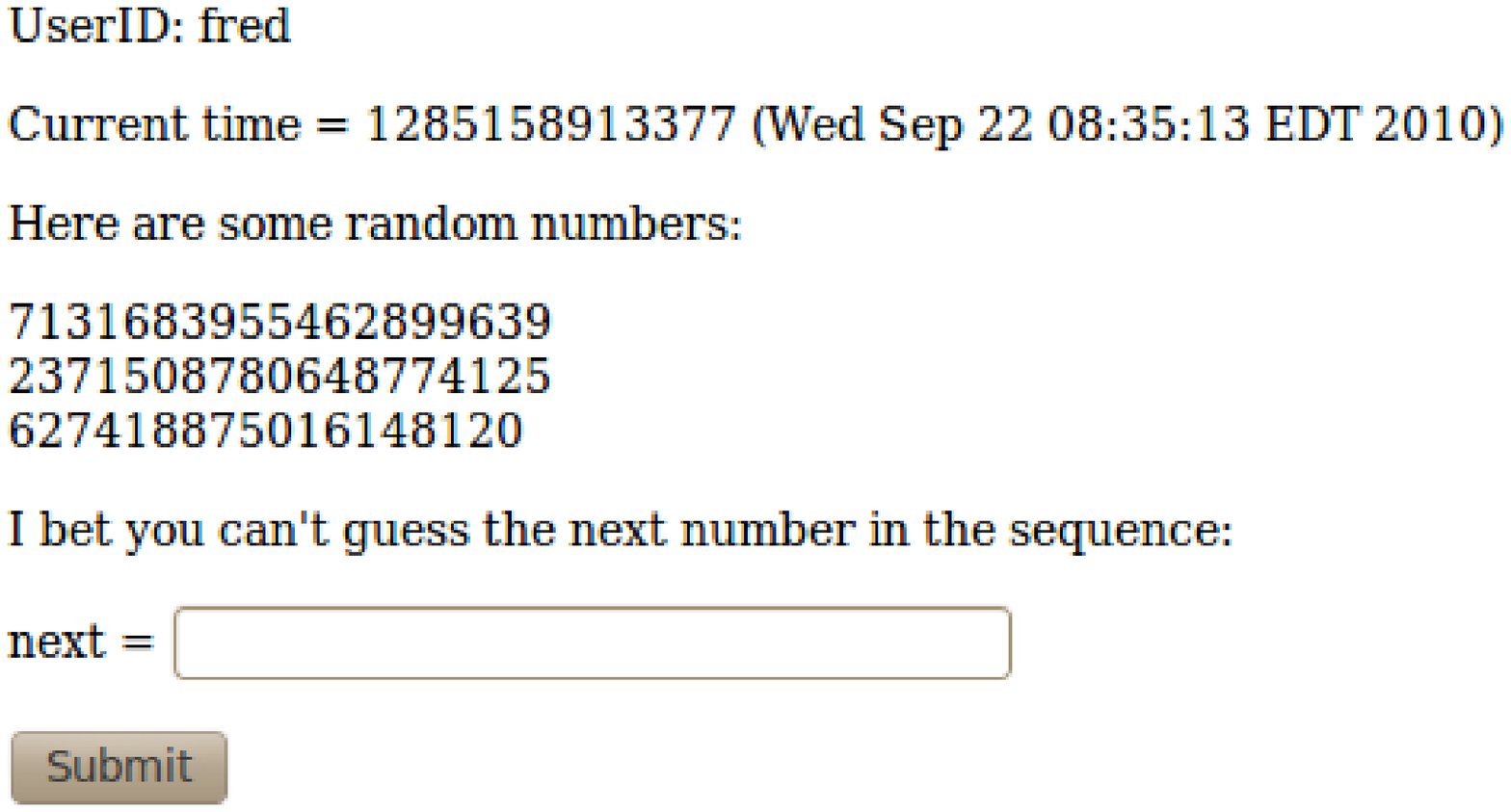}}
\caption{Dynamic Example Based on the Current Time}
\label{fig:RNG}
\end{center}
\end{figure}

\vspace{11pt}

Figure \ref{fig:RNG} shows a dynamic example
where the data depends on the date and time at which the student accesses the exercise.
The data is generated using Java's Random class, initialized using a value
close to the time of day in milliseconds.  Using a program to check times
near the one given, the student can reproduce the pseudo--random sequence
and generate the next value:

\begin{verbatim}
    UserID: fred

    Your answer is correct. You win!

    That was fun. Are you ready for a harder problem?

    Try this: I'll give you just one value from nextLong(), using an instance
    of Random initialized in a secret way, not related to the time of day.
    And I bet you can't guess the next number...<link to continue here>
\end{verbatim}

\begin{figure}[ht]
\begin{center}
\fbox{\includegraphics[scale=0.5,angle=0]{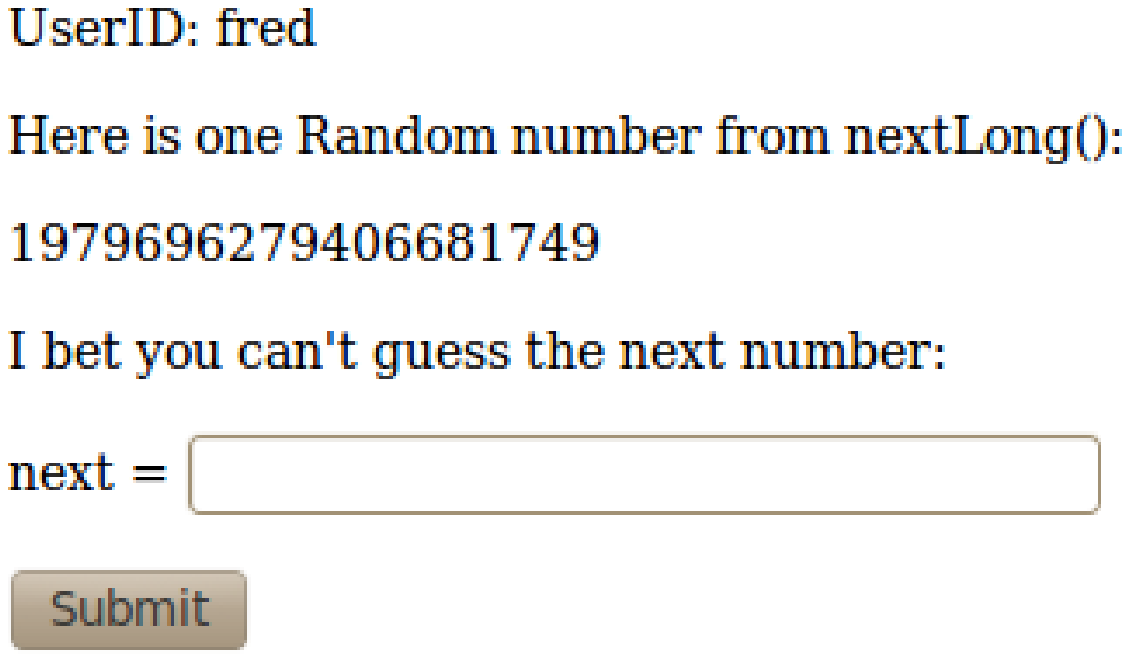}}
\caption{Dynamic Challenge Exercise}
\label{fig:RNG1}
\end{center}
\end{figure}

\vspace{11pt}

As shown above, when the correct answer is submitted,
the student is challenged to solve a harder problem.  If the student proceeds,
a new exercise is generated dynamically, 
based on Java's Random class initialized in an unpredictable manner,
as shown in Fig.\ \ref{fig:RNG1}.
If the student is able to solve this harder problem, they are congratulated:

\begin{verbatim}
    UserID: fred

    Your answer is correct.

    I give up! You are the master of pseudo-random numbers!
\end{verbatim}

\begin{figure}[ht]
\begin{center}
\fbox{\includegraphics[scale=0.5,angle=0]{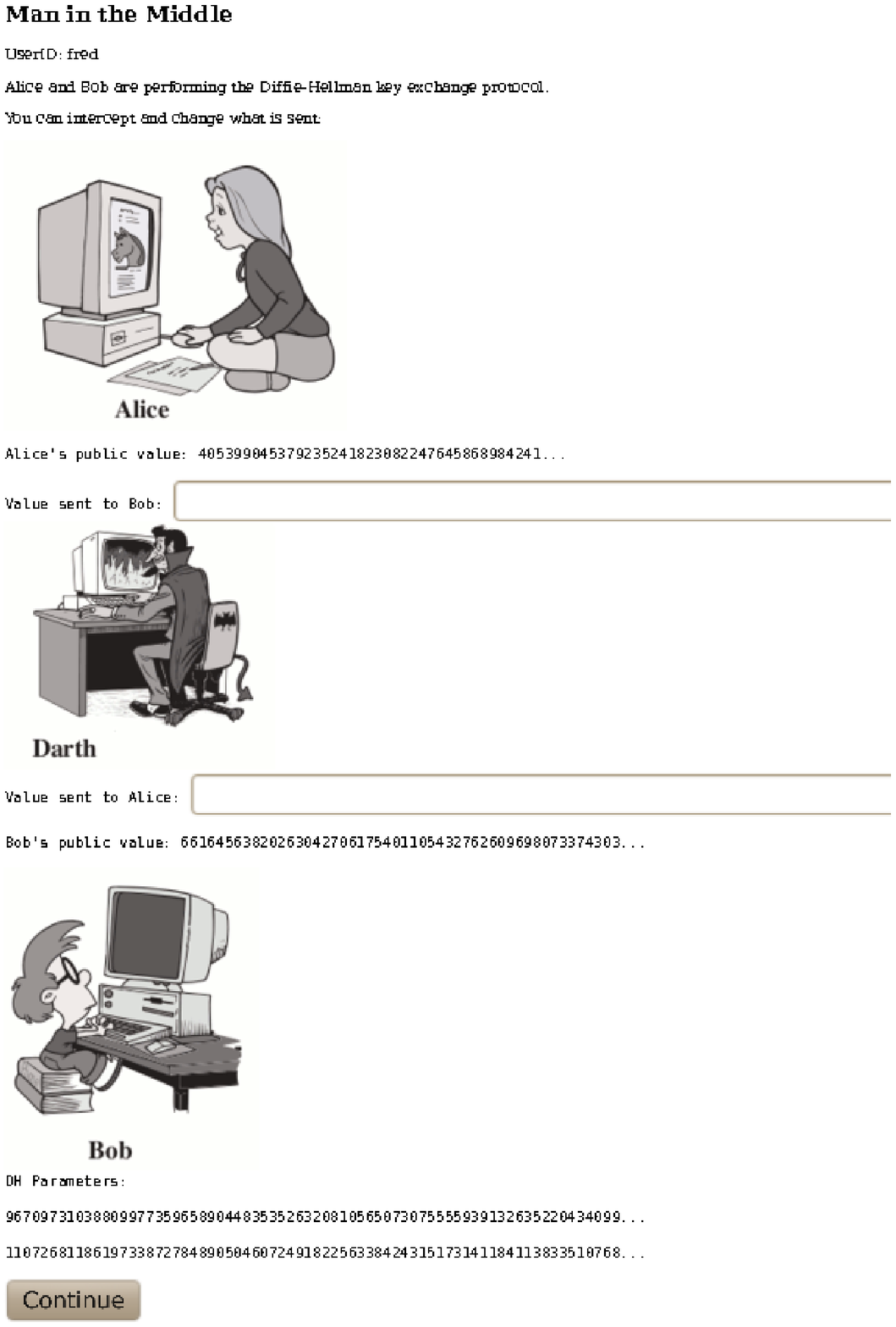}}
\caption{Man--in--the--middle Part 1}
\label{fig:MITM1}
\end{center}
\end{figure}

\begin{figure}[ht]
\begin{center}
\fbox{\includegraphics[scale=0.5,angle=0]{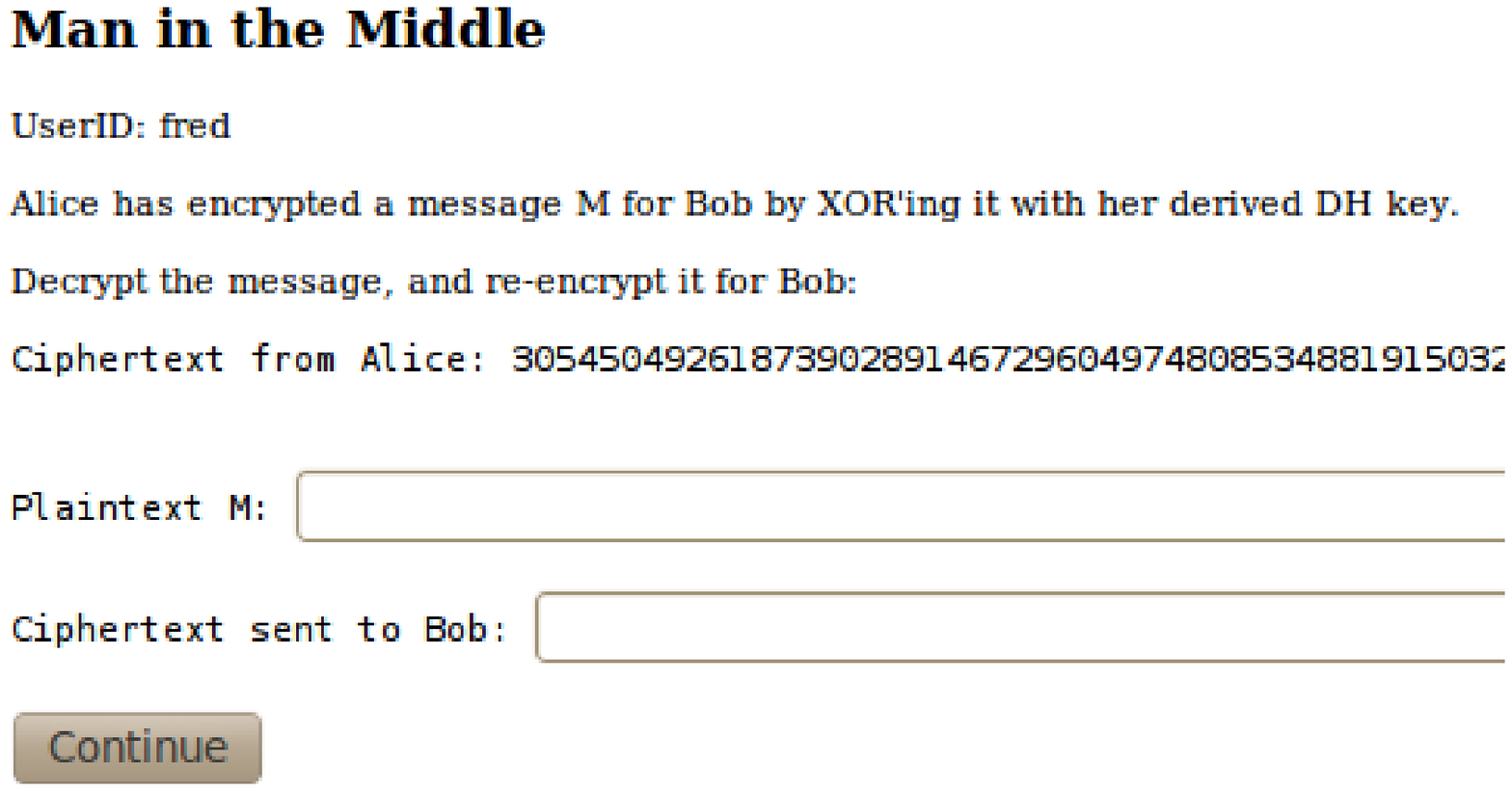}}
\caption{Man--in--the--middle Part 2}
\label{fig:MITM1-post}
\end{center}
\end{figure}

In the previous example the exercise was split into two parts, with the
second part representing a challenge which is presented only after the
first part is solved correctly.  It is also possible to split an exercise
into multiple parts based on logical aspects of the problem being solved.
An example of this is shown in Figs.\ \ref{fig:MITM1} and \ref{fig:MITM1-post},
where the student initiates a man--in--the--middle attack on the
Diffie--Hellman key agreement protocol in the first part,
and in the second part the attack continues by decrypting the secret communications.
The caricatures of Alice, Darth, and Bob in Fig.\ \ref{fig:MITM1}
are from Stallings \cite{ws5}.

\vspace{11pt}

If incorrect results are submitted, the student is informed about
which parts were wrong, and encouraged to try again:

\begin{verbatim}
    Man in the Middle

    UserID: fred

    Your value for M is wrong and your value for Cb is wrong. (checking...
    Ka=1 Kb=1) Ka and Kb are equal, that's non-standard! But Ka and/or Kb
    are trivial, that doesn't count. XTa 1 XTb 1

    Get back in the middle and try again.
\end{verbatim}

\vspace{22pt}
Real--time Dynamic Examples
\vspace{11pt}

For real--time dynamic exercises, students must submit their answers
within a small time window.  In the following examples the time window
is 60 seconds, i.e.\ the connection is dropped if the student does not
respond to any prompt within that time limit.
This is implemented on the server side using a simple socket timeout
option, and prevents failed server processes from accumulating
indefinitely.

\vspace{11pt}

In these examples, the student uses a telnet client to communicate with the
server and values are sent and received in plain text using hex.
In a more advanced course, the student may be required to write a
C or Java program to perform the network communications, sending and
receiving binary values directly.  In that case
the timeout would be set much lower, e.g.\ 10 seconds.

The first example demonstrates a one--way
authentication attempt where the student does not respond correctly to the challenge.
In a correct response, the student would use their personal course--assigned
password key to encrypt the challenge.  Seinfeld fans may note a similarity
to a certain soup kitchen episode in the authentication failure response:

\begin{verbatim}
    % telnet fog.misty.com
    Trying 198.137.254.19...
    Connected to fog.misty.com.
    Escape character is '^]'.
    UserID: fred
    Request #: 1
    Challenge: e9b24781e1fc0037
    Response: abc

    Authentication failed.  No fortune for you!
\end{verbatim}

The next example shows one--way authentication with a correct response
from the student.  In this case the student is rewarded with a random fortune.
Note that the challenges are generated randomly and will vary with each
connection attempt:

\begin{verbatim}
    UserID: fred
    Request #: 1
    Challenge: 88ce062ea5f139b6
    Response: ee47da0f6691bc60

    Authentication succeeded.  Here is your random fortune:

    "And it's my opinion, and that's only my opinion, you are a lunatic.
    Just because there are a few hundred other people sharing your lunacy
    with you does not make you any saner.  Doomed, eh?"

                    -- Oleg Kiselev,oleg@CS.UCLA.EDU
\end{verbatim}

The next level of the exercise uses two--way authentication, 
where the student challenges the server to encrypt a value. 
To ensure that the student decrypts the
server response and checks the result, the server chooses one of the
challenge bytes at random and changes it to a random value.  After decrypting,
the student must identify which byte was changed and return that as the
``check byte'':

\begin{verbatim}
    UserID: fred
    Request #: 2
    Challenge: 1c3ec315353599ac
    Response: e8f6b957e9d2b0ed
    Your challenge for me: aabbccdd00112233
    My response: bb738ef05d1497c9
    Check byte: cb

    Authentication succeeded.  Here is your random fortune:

    The face of war has never changed.  Surely it is more logical to heal
    than to kill.
            -- Surak of Vulcan, "The Savage Curtain", stardate 5906.5
\end{verbatim}

The final level of the exercise uses two--way authentication with confidentiality.
In this case, the random fortune
is encrypted using a key which depends on the student course--assigned password
and the two challenges:

\begin{verbatim}
    UserID: fred
    Request #: 3
    Challenge: 841012a067425b1a
    Response: aef79354a1b6cec7
    Your challenge for me: 0011223300112233
    My response: b41808205c531837
    Check byte: 74

    Authentication succeeded.  Here is your random fortune:

    KVvUeBXI/0kQ3A8TFa/G4zBlSkVPxLOOdgBIF0QGYdqu78eFR5vacYmqcIHp
    XQlRbQjVdH1f1Q5OdOh+mczG+uGUNXiPV98pJEQVOLVjmE7OSKrwx798oGyx
    530fkjYM2wb55qzF4khrJB8mu94qBdP9Qm8XdePI/HrIEkypqUU76ALrGs0+
    GWFBvHp5VQOJXwQGbTfqgoBzJQaMBp20WEYoXL5ZfNQ2nE9rM8ocVMb+nYIy
    1Jzn+SUgOAq9gZrQDEDpWxc0Lo/rEBlMcxH9Wpu2zXuwqQPE92nXmkOI+5Bw
    Tn8xqq5gDwbHq+zqdO2sOL7KeMyVSy0eiQ==
\end{verbatim}

When the student decrypts the message, the random
fortune and a ``user authentication code'' are displayed:

\begin{verbatim}
    To a Californian, the basic difference between the people and the
    pigeons in New York is that the pigeons don't crap on each other.
            -- From "East vs. West: The War Between the Coasts

    7144cd971a1c9d8b5d9ac68c4ae1eef1a206e781e1e47f73e2c34e93ed6cd06e
\end{verbatim}

\begin{figure}[ht]
\begin{center}
\fbox{\includegraphics[scale=0.5,angle=0]{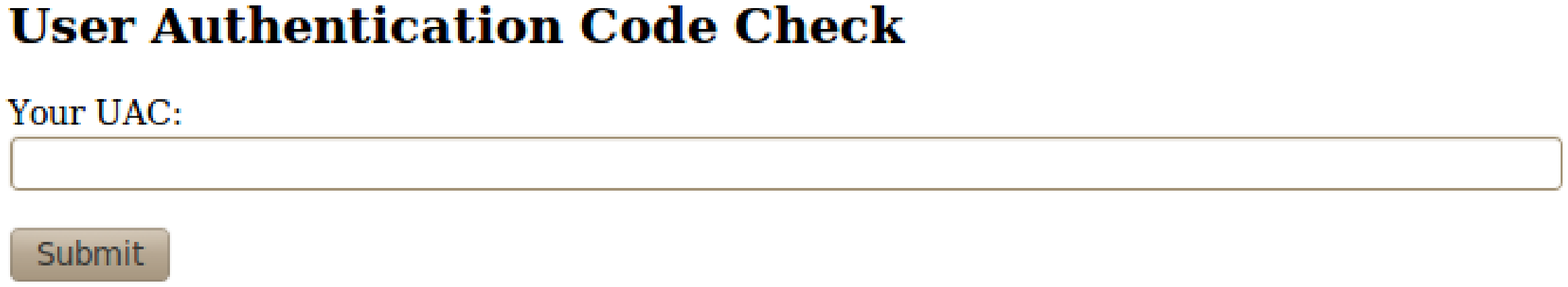}}
\caption{Submitting the User Authentication Code}
\label{fig:UAC}
\end{center}
\end{figure}

The user authentication code 
is generated pseudo--randomly, based on the student UserID, and is used
by the student to prove that they decrypted the message by submitting it
via a web form as shown in Fig.\ \ref{fig:UAC}.

\newpage
% \vspace{22pt}
Analyzing and Grading Student Results
\vspace{11pt}

Student results for each exercise are recorded in log files which 
can be easily processed in an automated fashion for grading.  
For example, the following results were generated from the log files
for an assignment which had four parts with each part worth 25 points:

\begin{verbatim}
    fred  25 25 25 25
    alice 25  5 25 15
    bob   25  0 25 25
    sam   25 25 10 25
    tony  25  0 25  0
    phil   0  0 25 25
    harry 25 15 25 25
    nancy 25  0 25 25
\end{verbatim}

Partial credit can be automatically computed for multi--part exercises
where the student only completed some of the parts correctly.
Some credit for effort can also be automatically assigned for a student who
did not supply a correct answer for an exercise but made many attempts
to do so.

\vspace{11pt}

Detailed results can be produced for each student and exercise,
showing when and how often the student attempted to solve the exercise,
including incorrect attempts, for example:
\begin{verbatim}
    25

    log count =  10

    Sun Apr  4 19:39:45 EDT 2010 fred You have 1 out of 3 parts correct.
    Fri Apr  9 22:58:24 EDT 2010 fred You have 2 out of 3 parts correct.
    Fri Apr  9 23:19:39 EDT 2010 fred You have 2 out of 3 parts correct.
    Fri Apr  9 23:26:14 EDT 2010 fred You have 2 out of 3 parts correct.
    Fri Apr  9 23:32:49 EDT 2010 fred You have 2 out of 3 parts correct.
    Sat Apr 10 00:20:24 EDT 2010 fred You have 2 out of 3 parts correct.
    Sat Apr 10 00:31:49 EDT 2010 fred You have 2 out of 3 parts correct.
    Sat Apr 10 00:38:52 EDT 2010 fred You have 2 out of 3 parts correct.
    Sat Apr 10 11:33:25 EDT 2010 fred You have 2 out of 3 parts correct.
    Sat Apr 10 18:37:42 EDT 2010 fred You have 3 out of 3 parts correct.
\end{verbatim}

This shows that fred had no trouble with parts 1 and 2 of the exercise,
obtaining the correct answers on the first try, but had some trouble
with part 3, finally supplying the correct answer after 7 or 8 failed
attempts.

\vspace{11pt}

Students are never penalized for incorrect attempts; in fact, they are encouraged to 
enter random junk to start with for an exercise just to see how the results
are processed, and this is generally demonstrated in class when a set of exercises
is first assigned.

\vspace{11pt}

In computer security courses the
students are also encouraged to examine the exercise interfaces closely and
try to ``break'' the system if they can, i.e.\ try to have a correct
response logged without actually supplying a correct answer.
So far, that has never happened, although maybe it did and was just
not detected.

\vspace{22pt}
Conclusion
\vspace{11pt}

The approach for engineering student exercises using the Internet
was demonstrated using
examples from computer security and cryptography courses.
For a given exercise, each student
receives the same problem, but with different data.  
This approach is applicable to any type of engineering exercise
where the correct answers are suitable to be checked automatically,
which includes numerical and computational types of exercises, and perhaps others.

\vspace{22pt}
Bibliography
\vspace{-44pt}

% \bibliography{asee}

\begin{thebibliography}{1}

\bibitem{ws5}
W.~Stallings, {\em Cryptography and Network Security: Principles and Practice,
  Fifth Edition}.
\newblock Prentice Hall, 2010.

\end{thebibliography}

\end{document}